\documentclass[a4paper,11pt]{article}
\usepackage[utf8]{inputenc}
\usepackage{verbatim,graphics,graphicx,color,slashed,amsmath,bm}
\usepackage[T1]{fontenc}
\usepackage{subfigure}
\usepackage[table,svgnames]{xcolor}
\usepackage{color}
\usepackage{jheppub}
\usepackage{amsfonts}
\usepackage{amssymb}
\usepackage{bbding}
\usepackage{appendix}
\usepackage{mhchem}
\usepackage{stmaryrd}
\usepackage{blkarray}
\usepackage[export]{adjustbox}
\graphicspath{ {./images/} }
\usepackage{bbold}
\usepackage{CJKutf8}
\usepackage{float}
\usepackage[T1]{fontenc}

\allowdisplaybreaks[3]

\newcommand{\be}{\begin{equation} \begin{aligned}}
\newcommand{\ee}{\end{aligned} \end{equation} }
\newcommand{\bea}{\begin{eqnarray}}
\newcommand{\eea}{\end{eqnarray}}

\usepackage[normalem]{ulem}

\newcommand{\mc}{\mathcal}
\newcommand{\bl}{\mathbf}
\newcommand{\bs}{\boldsymbol}

\RequirePackage[normalem]{ulem} 
\RequirePackage{color}\definecolor{RED}{rgb}{1,0,0}\definecolor{BLUE}{rgb}{0,0,1} 

\title{Positivity bounds at one-loop level: the Higgs sector}

\author[a,b]{Xu Li}

\affiliation[a]{Theoretical Physics Division, Institute of High Energy Physics, Chinese Academy of Sciences, Beijing 100049, China}
\affiliation[b]{School of Physical Sciences, University of Chinese Academy of Sciences, Beijing 100049, China}

\emailAdd{lixu96@ihep.ac.cn}

\abstract{
	In this paper, we promote the convex cone method of positive bounds from tree level to loop level. This method is general and can be applied to obtain leading $s^2$ positivity bounds on the forward scattering process in the standard model effective field theory. To obtain the loop level bounds, the original tree level bounds are modified by loop corrections, which involve low dimensional coefficients. New positivity bounds being valid at one loop level on  the four-Higgs scattering have been provided.  We study some specific ultraviolet models to check the validity of the new bounds. In addition, the renormalisation group effect on positivity is explored. We point out that as long as the new bounds are satisfied at the cutoff scale $\Lambda$,  they will also be satisfied at all scales {below $\Lambda$.} 
	}

\begin{document}
\maketitle
\flushbottom


\section{Introduction}
\label{sec1}
Due to the lack of direct evidence for new physics at the Large Hadron Collider (LHC), the method of standard model effective field theory (SMEFT) has become the framework to indirectly detect new physics and interpret data. The SMEFT supplements the SM with higher dimensional operators $O^{(n)}_i$,
\bea
	\mathcal{L}_{\mathrm{SMEFT}}&=&\mathcal{L}_{\mathrm{SM}}+\sum_i \frac{C_i^{(6)}}{\Lambda^2} O_i^{(6)}+\sum_i \frac{C_i^{(8)}}{\Lambda^4} O_i^{(8)}+\cdots\;,
	\label{eq:SMEFT}
\eea
where $C^{(n)}_i$ is the Wilson coefficient of operator $O^{(n)}_i$ of dimension $n$. No matter what the real UV theory is, we can always integrate out the heavy particles and match the UV theory onto SMEFT by standard matching procedure. The $C^{(n)}_i$ will store the information of the heavy particles in the UV theory. As long as we assume the UV completion satisfies the axiomatic principles, such as unitarity, analyticity and locality, the Wilson coefficients are constrained by the positivity bounds~\cite{Pham:1985cr, Ananthanarayan:1994hf,Adams:2006sv}. 
In recent years, positivity bounds have made a lot of progress  theoretically~\cite{deRham:2017avq, deRham:2017zjm, Arkani-Hamed:2020blm, Bellazzini:2020cot, Tolley:2020gtv, Caron-Huot:2020cmc, Sinha:2020win, Alberte:2020jsk,Zhang:2020jyn, Li:2021lpe,  Chiang:2021ziz, Caron-Huot:2021rmr, Grall:2021xxm, Bern:2021ppb, Alberte:2021dnj, Du:2021byy, Bellazzini:2021oaj, Chowdhury:2021ynh, Caron-Huot:2022ugt, Chiang:2022jep, Chiang:2022ltp, Haring:2022cyf,Huang:2022mdb,Remmen:2022orj},
and have been widely used in cosmology, field theory and particle physics~\cite{Distler:2006if, Manohar:2008tc, Bellazzini:2015cra, Bellazzini:2016xrt, Cheung:2016yqr, Bonifacio:2016wcb, deRham:2017imi, Bellazzini:2017fep, deRham:2018qqo, Bonifacio:2018vzv, Melville:2019wyy, deRham:2019ctd, Alberte:2019xfh, Herrero-Valea:2019hde, Chen:2019qvr, Alberte:2020bdz, Huang:2020nqy, Tokuda:2020mlf, Wang:2020xlt, Wang:2020jxr, Herrero-Valea:2020wxz, deRham:2021fpu, Traykova:2021hbr, Arkani-Hamed:2021ajd, Haldar:2021rri, Raman:2021pkf, Gopakumar:2021dvg,Chala:2021wpj, Zahed:2021fkp, Kundu:2021qpi, Davighi:2021osh, Davis:2021oce, Alvarez:2021kpq, Melville:2022ykg, Li:2022tcz, Henriksson:2022oeu, Albert:2022oes,Chen:2022nym,CarrilloGonzalez:2022fwg}. 

The leading positivity bounds for the $s^2$ terms in the forward ($t=0$) limit ($s,t$ being the Mandelstam variables) of the amplitude are phenomenologically relevant for the dim-8 coefficients at tree level. The most widely used positivity bounds are obtained by considering forward elastic scattering of two external states~\cite{Adams:2006sv,Zhang:2018shp,Bi:2019phv,deRham:2018qqo,Cheung:2016yqr,Bellazzini:2018paj,Remmen:2019cyz,Remmen:2020vts,Wang:2020jxr,Andriolo:2020lul,Remmen:2022orj,Freytsis:2022aho,Ghosh:2022qqq,deRham:2022sdl}. 
However, the elastic bounds are not always optimal. In some cases, the bounds obtained by using convex geometry method may be tighter than the elastic bounds~\cite{Zhang:2020jyn,Fuks:2020ujk,Yamashita:2020gtt,Gu:2020ldn,Li:2021lpe,Zhang:2021eeo,Li:2022tcz}. 
In the convex geometry approach, finding the positivity bounds is equivalent to identifying the extremal rays of the positivity cone. The convex cone is constructed by working out the generators of irreducible representations (irreps) from the Clebsch-Gordon (CG) coefficients. Therefore, as long as we know all the symmetries of the external states of the forward scattering, the relevant bounds can be achieved to constrain the Wilson coefficients.

The above bounds are acquired under the assumption of tree level scattering, thus may become invalid when we consider the modifications of the loop diagrams. {There have been some works study one-loop level positivity bounds in the generic EFT} in Refs.~\cite{Bellazzini:2020cot,Bellazzini:2021oaj,Arkani-Hamed:2020blm,Chen:2022nym}. {These literatures consider EFTs that contain only one particle who is a scaler or a goldstone boson, and they drive positvity bounds on coefficients of $s^n$ terms, with $n \geq 2$. However, to our best knowledge, the loop-level positivity bound for only $s^2$ terms in the SMEFT framework, which usually contains more than one particle, are still lacking. }
Besides, only for dim-8 coefficients {in the SMEFT}, some specific UV models {can} induce loop-matching coefficients that destroy existing tree level positivity bounds for four-Higgs (4-$H$) scattering. Even though dim-8 coefficients satisfy positivity bounds at the cutoff scale, they may be driven away from the positivity region by their renormalisation group (RG) running~\cite{Chala:2021wpj}. 

In the present paper, we study the generalization of positivity bounds at the one-loop level with the convex geometry method. This method is suit for any forward scattering, whether the initial state and the final state are the identical. We mainly focus on the Higgs sector, i.e. consider the forward scattering between 4-$H$ in SMEFT. Then, we carry out a complete one-loop calculation and collect all possible contributions that can provide $s^2$-terms to the amplitude. The effect of one-loop RGE is also included in our consideration. In addition, We will verify the correctness of the new positivity bounds with several special UV models as examples.

The paper is organized as follows. In section \ref{sec:formalism}, we review the formalism of the convex cone approach for positivity. In the process of obtaining the master formula, the tree level assumption is not adopted, so the framework can be used for the loop order. In section \ref{sec:treebound}, we apply the convex approach to 4-$H$ process at tree level to obtain the existing bounds that are widely known. The loop level corrections to positivity bounds as well as the several explicit examples of UV models are analyzed and discussed in section \ref{sec:loopbound}. Section~\ref{sec:RGE} explores RGEs effects of the dim-8 Wilson coefficients and the variation of bounds with the scale. We summarise and conclude in section~\ref{sec:summary}.

\section{Formalism}
\label{sec:formalism}
\begin{figure}[t]
	\centering  
	\subfigure[Amplitude]{
	\label{amp}
	\includegraphics[width=0.4\linewidth]{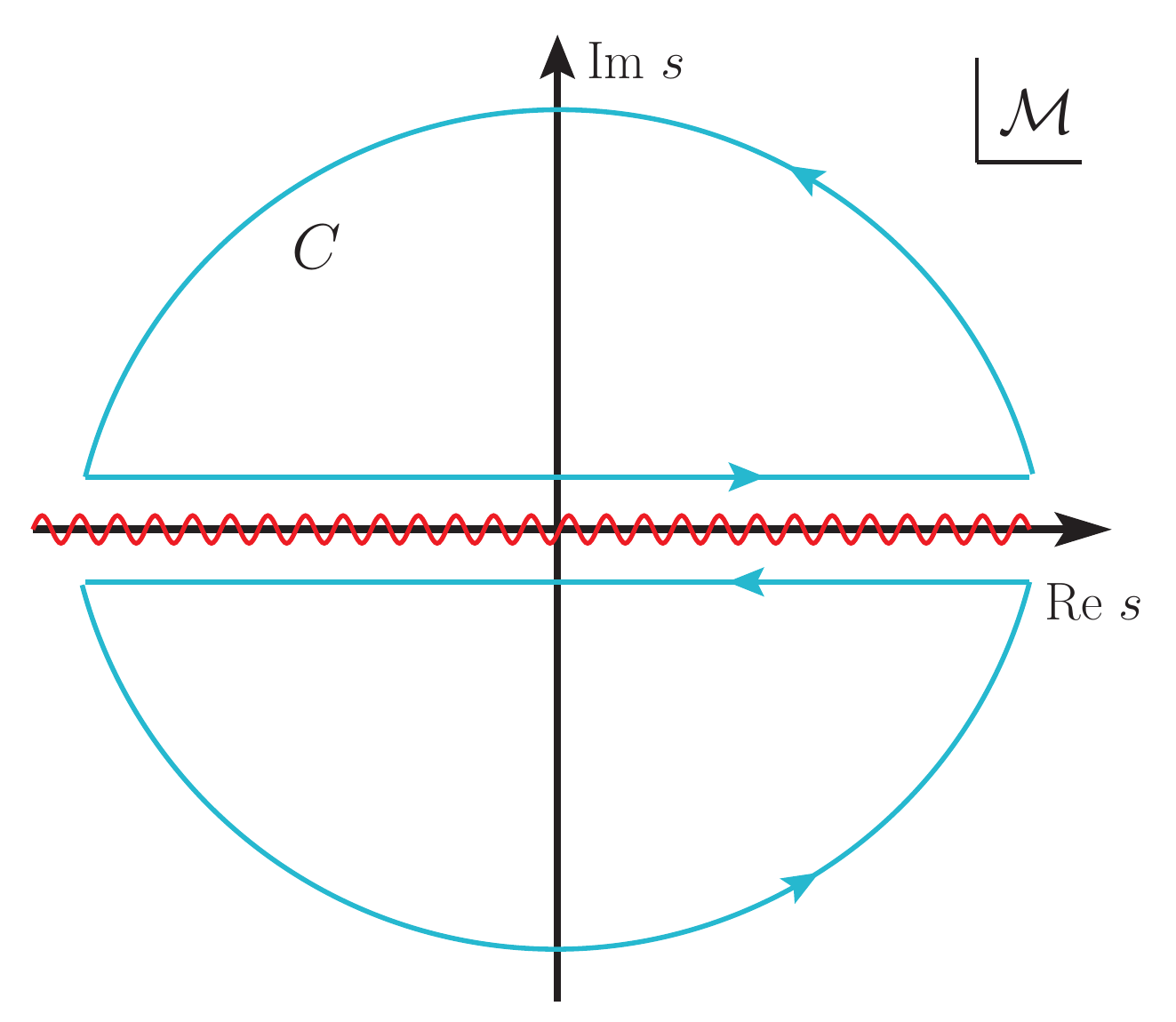}} 
	$\ \ \ $
	\subfigure[Subtracted amplitude]{
	\label{subamp}
	\includegraphics[width=0.4\linewidth]{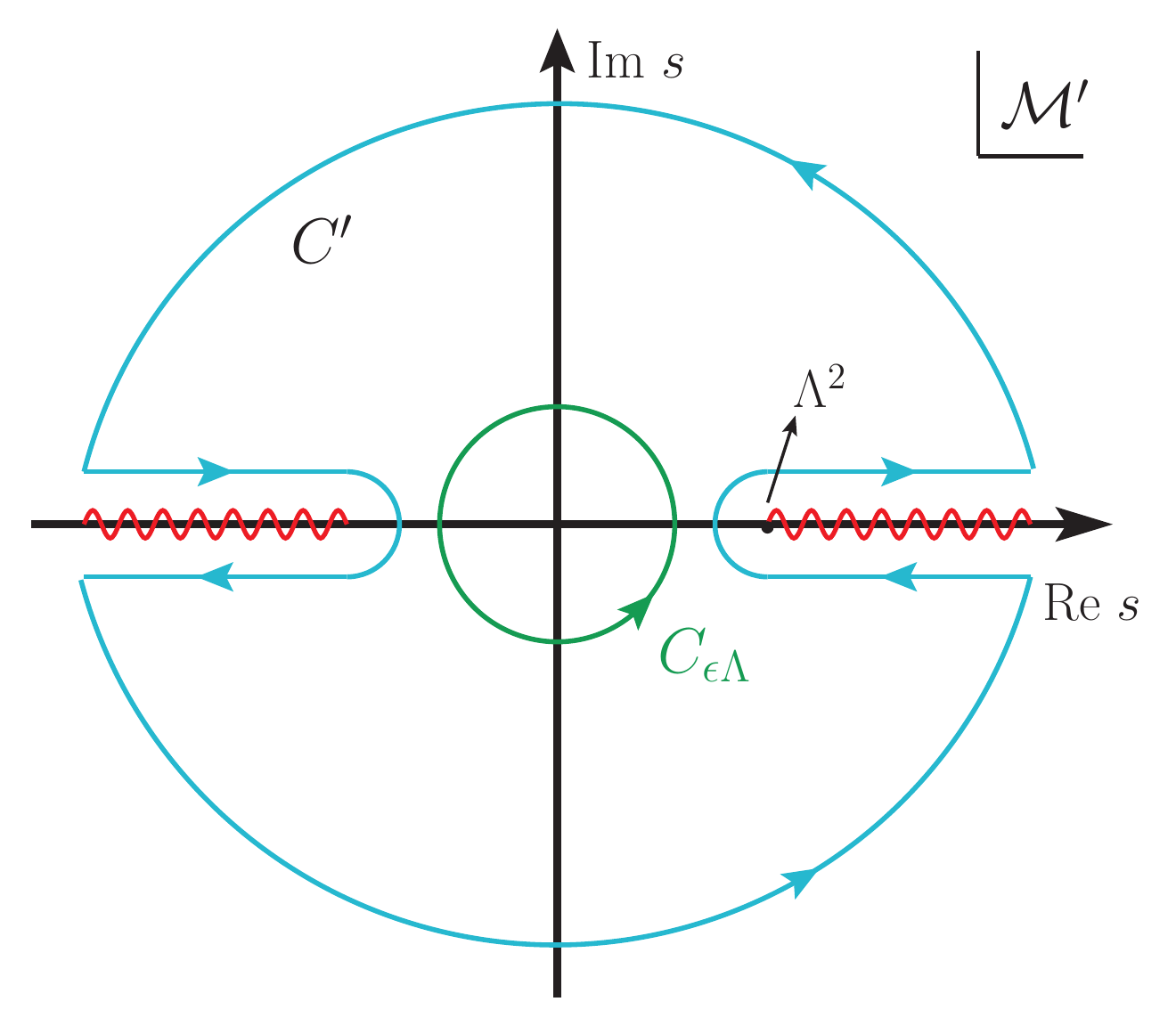}}
	\caption{The contour integration on the analytical $s$-plane of the amplitudes.}
	\label{contour}
\end{figure}

We first review the general analytic structure of the forward scattering amplitude up to one loop order. {For the fixed $t$(we focus on $t=0$), the analyticity of the scattering amplitude requires that} the amplitude is only an analytic function of $s$. On the complex-$s$ plane, the singularities of the amplitude include poles and branch cuts, which basically locate on the real axis. The pole corresponds to the mass of the state appearing in the propagator of the tree diagram. 
And the branch cut relates to the production of multi-particle states in the loop,  expands from the threshold of the production to infinity on the real axis. 
The branch cut is symmetric about the origin of the axis because the amplitude, as is implied by the crossing symmetry, is invariant under $s\leftrightarrow -s$. Therefore, when the multi-particle states are all massless, the branch cut can extend to the whole axis\footnote{For the non-analyticity of the massless theory at loop level, except for the massless cut, there can be IR singularities such as the soft or the collinear divergences. However, for the 4-$H$ processes we considered in Section~\ref{sec:loopbound}, such IR singularities do not appear, thus the forward limit is well-defined. }. 
Except for these singularities, the amplitude is analytical on the $s$-plane, see Fig.~\ref{amp}. For a general amplitude on $s$ that is away from these singularities, we have: 
\bea
M\left(s\right)=\oint_{C}{ds^\prime\frac{M\left(s^\prime\right)}{s^\prime-s}}
\label{eq:pole}
\eea
where the contour $C$ has been chosen to avoid the singularities of the amplitude, as shown by Fig.~\ref{amp}. Eq.(\ref{eq:pole}) works because the integrand only has a pole on $s'=s$ that stays inside the contour. In this paper, we are concerned with the leading positivity bounds on the $s^2$ terms of the amplitude. To extract the $s^2$-dependence from the amplitude, we define a quantify $f$ to be the second derivative of the amplitude to $s$, for the process $ij\to kl$, with $i,j,k,l$ are  particle indices,
\bea
f=\left. \frac{1}{2}\frac{d^2}{ds^2}M_{ij\to kl} \left(s, t=0 \right)\right|_{s\to 0}\;.
\eea
However, this formula has an underlying problem. The existence of singularities can make the function $f$ singular in the limit $s\to 0$. At the tree level, the low-energy region of $M_{ij\to kl}$ can be calculated by the EFT. In the massless limit, there is a pole at $s=0$ originating from the propagator of the SM particle.  When we calculate the loop diagrams in EFT, the $\log{(-s)}$ term is inevitable, and it is a branch cut extending from the origin to infinity on the real $s$-axis. In this situation, such a log-term is singular in the limit $s\to 0$. To address the $s\to 0$ singularity, previous literatures have considered several methods such as introducing a small mass $m$ for the internal propagator, or moving the contour from origin to the complex plane~\cite{Bellazzini:2020cot,Bellazzini:2021oaj,Arkani-Hamed:2020blm,Chen:2022nym}. In the present paper, we make use of the ``subtracted amplitude'' to remove such a singularity~\cite{deRham:2017zjm}.

\subsection{Subtracted Amplitude}
The full amplitude is defined by the contour $C$ in Fig~\ref{amp}, which consists of an integral on discontinuities along the entire real $s$ axis, and two semi-circles at infinity which vanish because the amplitude falls fast at infinity. We divide the integral region into the infrared (IR)-region and the UV-region with separation points at $s=\Lambda^2$ and $s=-\Lambda^2$, and subtract the IR-part of this integral from the amplitude. In this way, the amplitude after subtraction will not include any singularity in the low-energy region, i.e. low energy poles and low energy  branch cuts vanish in the amplitude. More precisely, 
we define the $M'$ as the subtracted amplitude:
\bea
M^\prime\left(s\right)&=&M\left(s\right)-\frac{1}{2\pi i}\int_{- \Lambda^2}^{+\Lambda^2}{ds^\prime\frac{{\rm Disc} M\left(s^\prime\right)}{s^\prime-s}\ }\nonumber \\
 &=&\frac{1}{2\pi i}\left(\int_{-\infty}^{- \Lambda^2}{ds^\prime}+\int_{ +\Lambda^2}^{\infty}{ds^\prime}\right)\frac{{\rm Disc} M\left(s^\prime\right)}{s^\prime-s}\nonumber \\
&=&\frac{1}{2\pi i}\oint_{C_{\epsilon\Lambda}}{ds^\prime\frac{M^\prime\left(s^\prime\right)}{s^\prime-s}}\;.
\label{eq:subAmp}
\eea
In the last line we have deformed the contour $C'$ to a small circle $C_{\epsilon \Lambda}$ by the Cauchy theorem, with $0<\epsilon<1$, see Fig.~\ref{subamp}.  {Due to the analyticity of the subtracted amplitude inside $|s|<\Lambda^2$, the theory will not change when we change the radius of the dispersion relation $\epsilon \Lambda$. The $\sqrt{s}\leq \epsilon \Lambda $ is where we probe the theory and the} EFT is valid inside $C_{\epsilon \Lambda}$. 
It can be seen from Eq.(\ref{eq:subAmp}) that
\begin{itemize}
	\item $M(s)$ contains singularities on the whole real $s$ axis, thus it equals to the integral on the discontinuities of $\frac{M\left(s^\prime\right)}{s^\prime-s}$ along the region $[-\infty,+\infty]$.
	\item $M^\prime\left(s\right)$ doesn't contain any discontinuity in $[-\Lambda^2,+\Lambda^2]$. We only need to integrate the interval region in which $|s|$ is greater than $\Lambda^2$.
\end{itemize}

It is not difficult to subtract the amplitude. We give an example for the subtraction scheme in appendix~\ref{appA}. The result is very simple. The single pole whose location is lower than $\Lambda^2$, such as $\frac{1}{s-m^2}$ with $m^2< \Lambda^2$, will be removed from in $M(s)$ directly. Besides, the log-term $\log{(-s)}$ appearing in the loop calculation transforms as
\bea
\log{(-s)}\to \log{(-s)} -\frac{1}{2\pi i}\int_{0}^{\Lambda^2}{ds^\prime\frac{{\rm Disc} \log{(-s^\prime)}}{s^\prime-s}\ } = \log{( \Lambda^2 -s)}
\eea
where the ${\rm Disc} \log{(-s)}=- 2\pi i$ have been used. This transformation can be understood as follows: the branch cut starting from the origin of real axis, after the subtraction, starts from $\Lambda^2$. 
{Nevertheless, to address the IR singularities, different prescriptions have been taken in previous literatures. One can introduce a small mass $m$ for the internal propagator as a regulator~\cite{Bellazzini:2020cot,Bellazzini:2021oaj,Arkani-Hamed:2020blm,Chen:2022nym}. However, the corresponding bounds may cause tricky problems when $m\to 0$. Alternatively, Ref~\cite{Arkani-Hamed:2020blm} defines the generalized effective couplings by moving the contour from origin to the complex plane. But under this circumstance, the tree-level coefficients are subdominant to terms that are generated from logarithms of loop calculations. In contrast, we will see that our  prescription only produces small loop corrections to the tree-level coefficients.}

Note the form of Eq.(\ref{eq:subAmp}) is similar to that of Eq.(\ref{eq:pole}). The difference is that now the contour is $C_{\epsilon\Lambda}$, in which the $M'(s)$ is analytic. 
The utilization of {the} subtracted amplitude can make sure the effectiveness of loop calculation, and does not affect the evaluation of the integral on the UV-region. After that, we only need to prove that the second derivative of the subtracted amplitude will be constrained by the positivity bounds.

\subsection{Second derivative of the {subtracted} amplitude}
Directly taking the second derivative of Eq.(\ref{eq:subAmp}) to $s$ we obtain
\bea
f'&\equiv& \left.\frac{1}{2}\frac{d^2}{ds^2} M^\prime\left(s\right)\right|_{s\to 0}= \frac{1}{2\pi i}\oint_{C_{\epsilon\Lambda}}{ds^\prime\frac{M^\prime\left(s^\prime\right)}{\left(s^\prime\right)^3}} \nonumber \\ 
&=&\frac{1}{2\pi i}\left(\int_{-\infty}^{-(\epsilon \Lambda)^2}{ds^\prime}+\int_{  + (\epsilon \Lambda)^2}^{\infty}{ds^\prime}\right) \frac{{\rm Disc} M^\prime\left(s^\prime\right)}{\left(s^\prime\right)^3} 
\nonumber \\ 
&=&\frac{1}{2\pi i}\left[\int_{(\epsilon \Lambda)^2+m^2}^{\infty}\frac{du\ {\rm Disc} M^\prime\left(m^2-u\right)}{\left(m^2-u\right)^3}+\int_{(\epsilon \Lambda)^2}^{\infty}\frac{ds^\prime\ {\rm Disc} M^\prime\left(s^\prime\right)}{\left(s^\prime\right)^3}\right]\;,
\label{eq:secDe1}
\eea
where $m^2$ is the sum of the squared masses of external states, and variable is changed in last line: $s^\prime=m^2-u$. Utilizing the crossing symmetry, $M_{ij\rightarrow k l}\left(s\right)=M_{i\bar{l}\rightarrow k\bar{j}}(m^2-s)\equiv \overline{M}_{ij\to kl}(m^2-s)$, where $\overline{M}$ is crossing of $M$, the ${\rm Disc} M\left(m^2-s\right)$ becomes
\bea
{\rm Disc} M\left(m^2-s\right)&\equiv&M\left(m^2-s+i\epsilon\right)-M\left(m^2-s-i\epsilon\right) \nonumber \\ 
&=&\overline{M}\left(s-i\epsilon\right)-\overline{M}\left(s+i\epsilon\right) \nonumber \\ 
&\equiv&-{\rm Disc}\overline{M}(s) \;.
\eea
After that, taking the massless limit we can rewrite Eq.(\ref{eq:secDe1}) as
\bea
f'
&=&\frac{1}{2\pi i}\int_{\left(\epsilon\Lambda\right)^2}^{\infty}\frac{ds^\prime}{\left(s^\prime\right)^3}\left[{\rm Disc} M^\prime\left(s^\prime\right)+{\rm Disc}\overline{M}^\prime\left(s^\prime\right)\right]\;.
\label{eq:secDe2}
\eea
By the Schwartz reflection theorem $M_{ij\rightarrow k l}\left(s+i\epsilon\right)=M_{kl\rightarrow i j}^\ast\left(s-i\epsilon\right)$, and the generalized optical theorem, the discontinuity of the amplitude is
\bea
{\rm Disc} M \left(s\right)&=&M^{}_{ij\rightarrow k l}\left(s+i\epsilon\right)-M_{ij\rightarrow k l}\left(s-i\epsilon\right)\\ 
&=& M^{}_{ij\rightarrow k l}\left(s+i\epsilon\right)-M_{kl\rightarrow i j}^\ast\left(s+i\epsilon\right)\\ 
&=&i\sum_{X}\int{d\Pi_XM^{}_{ij\rightarrow X}\left(s+i\epsilon\right)M_{kl\rightarrow X}^\ast(s+i\epsilon)}\;.
\eea
Finally, we reach the master formula as follow:
\bea
\left.\frac{1}{2}\frac{d^2M^\prime\left(s,t=0\right)}{ds^2}\right|_{s\to 0}= \sum_{X}\int_{(\epsilon \Lambda)^2}^{\infty}{\frac{ds^\prime}{2\pi\left(s^\prime\right)^3}\int  d\Pi_X\left[M^{}_{ij\rightarrow X}M_{kl\rightarrow X}^\ast+\left(j\leftrightarrow l\right)\right]}\;.
\label{eq:master}
\eea
{It should be emphasized that the traditional bound $f\geq 0$ can also be corrected at the loop level, because residues of the IR singularities are affected by loops}. {One may be curious for what is the difference between the traditional bound $f\geq 0$ and the subtracted bound $f'\geq 0$. The answer is that the subtracted bound will be physically stronger than the traditional bound \cite{Bellazzini:2016xrt,deRham:2017zjm,deRham:2017imi}. For example, we consider the forward scattering for identical particles. The traditional bound $f\geq 0$ is acquired by taking different treatments to ``regulate'' singularities in the IR region in the  EFT~\cite{Bellazzini:2020cot,Bellazzini:2021oaj,Arkani-Hamed:2020blm,Chen:2022nym}.  
By contrast, this paper directly subtract the amplitude by these singularities, i.e. $f'=f-f_{\rm low}$, with $f_{\rm low}=\int^{\Lambda^2}_0ds'\left[{\rm Im}{M}(s')/s'^3\right]\geq 0$, because the imaginary part of the forward amplitude is positive. We stress that the subtracted bound $f'\geq 0$ is stronger since such a positive condition implies $f\geq f_{\rm low}$, which is a tighter constraint than $f\geq 0$.

With similar argument, the
terms with $s$-dependence being higher than 2 in the IR amplitude  can be dropped.  For the quantity $f$, 
dropping these terms is safe, since $d^2(s^n \log(s))/ds^2\to 0$ as $s\to 0$ for $n>2$. However, one may find that these higher-$s$ contributions can enter the dispersion relation through $f_{\rm low}$ because they all contribute to the discontinuity.
Fortunately, we will prove that  higher-$s$ effects can only provide negative contributions to the bound $f'$, and neglecting them just strengthen the bound. In this sense, we are working in a truncated EFT whose amplitude is kept up to $\mc{O}(s^2)$.  
More explicitly, we divide the full EFT amplitude in the IR region into a truncated part and  a higher-$s$ part, $ {M}= {M}_{\rm truncated}+ {M}_{\text{higher-}s}$. The positivity condition becomes
\bea
&&f-\frac{2}{\pi}\int^{\Lambda^2}_{0}ds'\frac{{\rm Im} {M}_{\rm truncated}(s')}{s'^3}-\frac{2}{\pi}\int^{\Lambda^2}_{0}ds'\frac{{\rm Im} {M}_{\text{higher-}s}(s')}{s'^3} \nonumber \\
&&=
\sum_{X}\int_{(\epsilon \Lambda)^2}^{\infty}{\frac{ds^\prime}{\pi\left(s^\prime\right)^3}\int  d\Pi_X\left(M^{}_{ij\rightarrow X}M_{ij\rightarrow X}^\ast\right)} \geq 0\;.\nonumber
\eea
Note that the contribution from the higher-$s$ part to the l.h.s is negative because its integral is positive.  Removing this contribution from the equation will only strengthen the inequality. Therefore, it's safe for us to neglect the higher-$s$ amplitude, as we will do in the following content.
}

\subsection{Group theory for positivity}
\label{sec:group}
For the forward scattering $ij\to kl$ with $i,j,k,l$ being external states, we assume each state belongs to an irrep $\bl{r}$ of a symmetry groups. Two-particle states such as $\left|ij \right\rangle $ are the direct production of two irreps $\mathbf{r}_i$ and $\mathbf{r}_j$. They can be decomposed into irreps: $\mathbf{r}_i \otimes \mathbf{r}_j = \underset{\alpha} { \oplus}\ \bf{r}_\alpha $, where $\alpha$ is the index for inequivalent irreps. For example, in the 4-$H$ case where $H$ belongs to $\bf{2}$ of $SU(2)_{\rm L}$ symmetry, the decomposition is: $\bf{2} \otimes \bf{2} = \bf{1} \oplus \bf{3}$. 

The intermediate state $\left|X \right\rangle$ can be the one-particle extension of SM at tree level, or be two-particle states at one-loop level. In either case, we can decompose the  $\left|X \right\rangle$ into a sum of irreps, i.e. $\left|X \right\rangle = \sum_{\bl{r}_\alpha} C^{\bl{r}_\alpha}_X \left|\bl{r}_\alpha \right\rangle$. 
If $X$ is a one-particle state, it will be a linear combination of several irreps in which $C^{\bl{r}_\alpha}_X$'s  are arbitrary numbers. Provided that $X$ is a two-particle state $\left|ab\right\rangle$, $C^{\bl{r}_\alpha}_{a,b}$ denotes the CG coefficients which relates the direct product of $\bl{r}_a$ and $\bl{r}_b$ with $\bl{r}_\alpha$, meanwhile the sum is a direct sum. According to this view, it is sufficient to sum over all possible irreps in Eq.(\ref{eq:master}), in which the summation becomes:
\bea
\sum_{\bl{r}_\alpha}{M^{}_{ij\rightarrow \bl{r}_\alpha} M_{kl\rightarrow \bl{r}_\alpha}^\ast}&=&\sum_{\bl{r}_\alpha}{ \langle kl | \mc{M}^\dag  | \bl{r}_\alpha \rangle \left\langle \bl{r}_\alpha\middle| \mc{M}\middle| ij\right\rangle}\nonumber \\
&=&\sum_{\bl{r}_\alpha}\sum_{\bl{r}_{\rm in}, \bl{r}_{\rm out} }{\left\langle kl \middle|\bl{r}_{\rm in}\right\rangle  \langle \bl{r}_{\rm in} | \mc{M}^\dag  | \bl{r}_\alpha \rangle \langle \bl{r}_\alpha | \mc{M} |\bl{r}_{\rm out} \rangle\left\langle \bl{r}_{\rm out} \middle| ij \right\rangle}\nonumber \\
&=&\sum_{\bl{r}_\alpha} {C_{i,j}^{\bl{r},\alpha} C_{k,l}^{\bl{r},\alpha\ast} \left|M_{\bl{r}_\alpha\to\bl{r}_\alpha}\right|^2 } \;,
\label{eq:group}  
\eea
where complete sets of irreps $\bl{r}_{\rm in},\bl{r}_{\rm out}$ are inserted. The selection rule for scattering amplitude among different irreps $M_{\bl{r}_{\alpha}\to \bl{r}_{\beta} }(s,t)= \delta_{\alpha \beta} M_{\bl{r}_\alpha\to\bl{r}_\alpha}(s,t)$, which is guaranteed by the Wigner-Eckart theorem \cite{Bellazzini:2014waa}, has been utilized. 
Substituting Eq.(\ref{eq:group}) into Eq.(\ref{eq:master}) results in 
\bea
f'_{ijkl}=\left.\frac{1}{2}\frac{d^2M_{ij\to kl}^\prime\left(s,t=0\right)}{ds^2}\right|_{s\to 0}= \int_{(\epsilon \Lambda)^2}^{\infty}{\frac{ds^\prime}{2\pi\left(s^\prime\right)^3} \sum_{\bl{r}_\alpha} {\mc{G}_{\bl{r}_\alpha}^{i(j|k|l)} \int  d\Pi_{\bl{r}_\alpha} \left| M_{\bl{r}_\alpha\to\bl{r}_\alpha}\right|^2 }}\;.
\label{eq:finalEQ}
\eea
where we have defined $\mc{G}_{\bl{r}}^{ijkl}= \sum_{\alpha}C_{i,j}^{\bl{r},\alpha} C_{k,l}^{\bl{r},\alpha\ast}$, and $\mc{G}_{\bl{r} }^{i(j|k|l)}\equiv\mc{G}_{\bl{r} }^{ijkl}+\mc{G}_{\bl{r} }^{ilkj}$. In Eq.(\ref{eq:finalEQ}), the phase space integral of $|M|^2$ gives a cross section, thus is positive. Therefore, the $f'_{ijkl}$ is a positive sum of $\mc{G}_{\bl{r} }^{i(j|k|l)}$, or from the geometric perspective, a convex hull formed by various 
$\mc{G}_{\bl{r} }^{i(j|k|l)}$. For this reason, $\mc{G}_{\bl{r}}$ is referred to as a ``generator'' (for a cone) \cite{Zhang:2021eeo}. The $f'_{ijkl}$ in the low-energy region can be calculated in the EFT framework, either at tree order or loop order, thus is represented as a function of Wilson coefficients.  From a geometric point of view, the $\bs{f}'$ that is viewed as a vector is constrained inside a convex cone. Normal vectors $\bs{n}$'s of the cone have positive contractions with $\bs{f}'$, i.e. $\bs{n}\cdot \bs{f}'\geq 0$. More precisely, these inequalities are the positivity bounds for the given function $\bs{f}'$ of Wilson coefficients.  
To understand the structure of the cone and acquire it's normal vectors, we need to explore the right hand side (r.h.s) of Eq.(\ref{eq:finalEQ}) either at tree level or loop level.

\section{Tree level positivity}
\label{sec:treebound}
Positivity bounds for dim-8 coefficients of SMEFT will be derived at tree level. The function $f'_{ijkl}$ can be viewed as a matrix with $(ij,kl)$ being the indices, whose element $f'[ij,kl]$ indicates the value of $\frac{d^2}{2ds^2}M'(ij\to kl)$. The general structure of this amplitude matrix, in SMEFT, can be represented by the structure of dim-8 operators. There are three dim-8 operators that are relevant for 4-$H$ scattering at tree level~\cite{Li:2020gnx,Murphy:2020rsh,AccettulliHuber:2021uoa}:
\bea
\mc{O}^{(8)}_{H1} &=&  (D_\mu H^\dag D_\nu H)(D^\nu H^\dag D^\mu H) \nonumber  \\
\mc{O}^{(8)}_{H2} &=& (D_\mu H^\dag D_\nu H)(D^\mu H^\dag D^\nu H) \nonumber  \\ 
\mc{O}^{(8)}_{H3} &=& (D_\mu H^\dag D^\mu H)(D_\nu H^\dag D^\nu H)
\label{eq:Odim8}
\eea
Each operator $\mc{O}^{(8)}_{Hi}$ contributes to $\frac{d^2}{2ds^2}M'(ij\to kl)$, leading to an amplitude matrix that is characterized by the $\bl{m}_i$. Due to the completeness of the operator set, any amplitude matrix $f'_{ijkl}$ can be represented as a linear combination of the $\bl{m}_i$'s. In general, the $f'_{ijkl}$ lives in a linear matrix space spanned by three matrices $(\bl{m}_1,\bl{m}_2,\bl{m}_3)$. Such a space is referred to as the ``amplitude space'', and $f'_{ijkl}$ can be viewed as a vector $\bs{f}'$ inside the space. As we have mentioned, the decomposition of the direct product for 4-$H$ process is $\bf{2} \otimes \bf{2} = \bf{1} \oplus \bf{3}$. Provided the CG coefficients for various irreps, generators are given by Refs.~\cite{Zhang:2020jyn,Zhang:2021eeo} as
\begin{alignat}{3}
	\mc{G}_\mathbf{1}&=(1,0,-1),\quad \quad& \mc{G}_{\mathbf{1} S}&=(0,0,2),\quad \quad& \mc{G}_{\mathbf{1} A}&=(-2,2,0), \nonumber \\
	\mc{G}_{\mathbf{3}}&=(0,1,0), & \mc{G}_{\mathbf{3} S}&=(4,0,-2), & \mc{G}_{\mathbf{3}_A}&=(2,2,-4).
	\label{eq:generators}
\end{alignat}
The convex cone formed by these generators gives rise to out normal vectors as follows:
\bea
\bs{n}_1=(0,1,0),\quad \bs{n}_2=(1,1,0), \quad \bs{n}_3=(1,1,1)\;.
\label{eq:normalV}
\eea 
As is indicated in the Section~\ref{sec:group}, any amplitude $\bs{f}'$ is constrained inside the cone by these vectors, because of $\bs{n}_i \cdot \bs{f}'\geq 0$. At the tree level, only the three operators in Eq.(\ref{eq:Odim8}) can enter the 4-$H$ process and provide $s^2$ terms, i.e. $\mathcal{L}_{\mathrm{int}}=\sum_{i=1}^3 C_{Hi}^{(8)}O_{Hi}^{(8)}/\Lambda^4 $. Under these interactions, the amplitude matrix $\bs{f}'$ is given by
\bea
\Lambda^4 \bs{f}' &=& C_{H1}^{(8) }\ \bl{m}_1 + C_{H2}^{(8) }\ \bl{m}_2 + C_{H3}^{(8) }\ \bl{m}_3\;,
\eea
which can also be represented as the form of a vector $\vec{C}=(C_{H1}^{(8) },C_{H2}^{(8) },C_{H3}^{(8) })$.  The normal vectors given in Eq.(\ref{eq:normalV}) simply produce positivity bounds as
\bea
C_{H2}^{(8) } \geq 0,\quad C_{H1}^{(8)} +C_{H2}^{(8) } \geq 0,\quad C_{H1}^{(8) }+ C_{H2}^{(8) } + C_{H3}^{(8)} \geq 0 \;,
\label{eq:tree-bound}
\eea
which are first given in Ref.~\cite{Remmen:2019cyz} using the elastic approach, and are reproduced in Refs.~\cite{Zhang:2020jyn,Zhang:2021eeo} based on the convex cone approach. Eq.(\ref{eq:tree-bound}) is valid at tree order, but is not necessarily correct at one-loop order.

\section{Loop level positivity}
\label{sec:loopbound}
\subsection{Analytical bounds}
At the loop level in the SMEFT, not only dim-8 operators, but also dim-4 and dim-6 operators shall enter the calculation of $\bs{f}'$. 
At dim-4, the interaction is $-\lambda (H^\dag H)^2$.  And there are two operators that are directly relevant for $HH\to HH$ forward scattering at dim-6~\cite{Grzadkowski:2010es},
\be
		\mc{O}^{(6)}_{H1} &= \partial_\mu ( H^\dag  H) \partial^\mu ( H^\dag  H)\;,  \\
		\mc{O}^{(6)}_{H2} &= (D_\mu H^\dag H)(H^\dag D^\mu H) \;.
		\label{eq:dim6-H4}
\ee
The dim-6 operators that can't appear at tree diagram but take part in the loop diagrams are divided into two types:	
\begin{itemize}
	\item $\psi^2 H^2 D$
\begin{alignat}{2}
	\mathcal{O}_{H \psi_R} &= (H^{\dagger} i \overleftrightarrow{D}_\mu H )\left(\overline{\psi_R} \gamma^\mu \psi_R\right)\;, \quad & \mathcal{O}_{H u d} &= (\tilde{H}^\dag i D_\mu H )\left(\overline{u} \gamma^\mu d\right)+\text{h.c. } \nonumber \\
	\mathcal{O}_{H \psi_L}^{(1)} &= (H^{\dagger} i \overleftrightarrow{D}_\mu H )\left(\overline{\psi_L} \gamma^\mu \psi_L\right)\;, \quad & 
	\mathcal{O}_{H \psi_L}^{(3)} &= (H^{\dagger} i \overleftrightarrow{D}_\mu^I H )\left(\overline{\psi_L} \gamma^\mu \sigma_I \psi_L\right)\;,  
	\label{eq:dim6-Hpsi}
\end{alignat}
\item $X^2 H^2$
\begin{alignat}{2}
	 \mathcal{O}_{H V} &= H^{\dagger} H V_{\mu\nu} V^{\mu\nu}\;, \quad &
	 \mathcal{O}_{H \widetilde{V}} &= H^{\dagger} H \widetilde{V}_{\mu\nu} V^{\mu\nu}\;,  \nonumber \\
	 \mathcal{O}_{H WB} &= H^{\dagger}\sigma^I H W^I_{\mu\nu} B^{\mu\nu}\;,  \quad &
	 \mathcal{O}_{H \widetilde{W} B} &=  H^{\dagger}\sigma^I H \widetilde{W}^I_{\mu\nu} B^{\mu\nu}\;, 
	\label{eq:dim6-HV}
\end{alignat}
\end{itemize}
where $\overleftrightarrow{D}_\mu\equiv\overrightarrow{D}_\mu-\overleftarrow{D}_\mu$, and $\tilde{H}=\epsilon H^\ast$, $\sigma^I$ are the Pauli matrices, $V=(B,W^I,G^A)$, and $\widetilde{V}_{\mu\nu}=\frac{1}{2} \varepsilon_{\mu\nu\rho\tau} V^{\rho\tau} (\varepsilon_{0123}=+1)$. However, $X^2 H^2$-type operators can't be realized by tree diagram 
in any UV theory{, i.e. they are only obtained by UV loops. Single insertion of these operators into IR loops is suppressed by 2-loop  factor $1/(16\pi^2)^2$, while double insertion will be suppressed by 3-loop factor,}  thus they are not considered in our later discussion.

\begin{figure}[t]
	\centering  
	\includegraphics[width=0.8\linewidth]{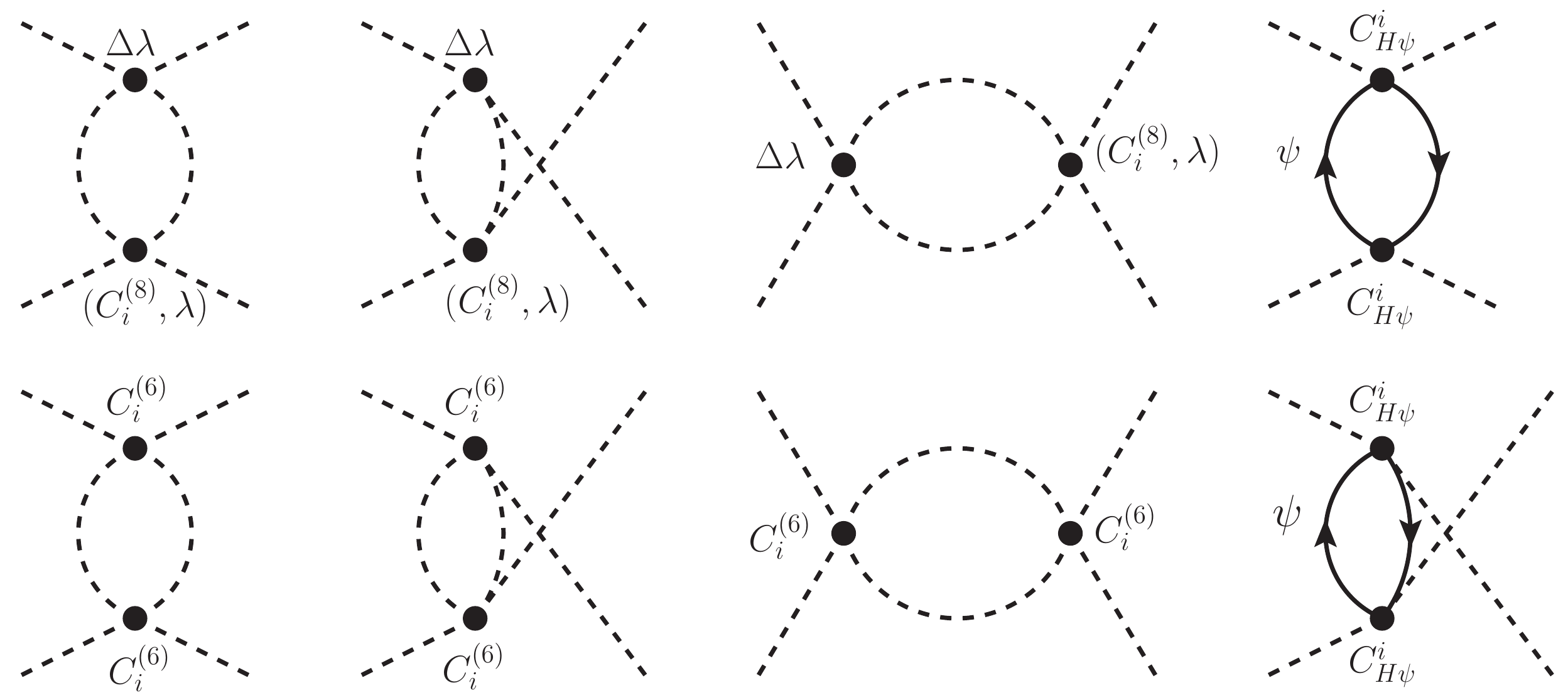}
	\caption{loop diagrams of the 4-$H$ process in EFT, the external legs are Higgs doublet, $\Delta \lambda$ is the threshold correction to SM coupling $\lambda$, meanwhile $C_{Hi}^{(8)},C_{Hi}^{(6)},C_{H\psi}^i$ indicate the one insertion of operators in Eq.(\ref{eq:Odim8}), Eq.(\ref{eq:dim6-H4}) and Eq.(\ref{eq:dim6-Hpsi}) respectively, the $\psi$ is the SM fermion $\psi_L$ or $\psi_R$.}
	\label{fig:EFTdiagrams}
\end{figure}

The one-loop diagrams in the SMEFT are shown in Fig.~\ref{fig:EFTdiagrams}. The double insertions of dim-6 operators, or one dim-8 term plus one dim-4 term, can arise in loop diagrams. For simplicity, we have neglected loop diagrams consisting of couplings of SM, i.e. have taken the $g_{\rm SM}\to 0$ limit\footnote{The rationality of this can be seen through Eq.\eqref{eq:master}, whose l.h.s is calculated in the SMEFT and the r.h.s is studied in the UV theory. The $g_{\rm SM}\to 0$ limit can be taken on both sides, which indicates that only tree diagrams and loop diagrams involving UV couplings are considered. Limiting the scope of diagrams considered  does not affect the positiveness of the $f'$, which is guaranteed by the cutting rule.}. Therefore, the loop amplitude for 4-$H$ scattering up to  $\mc{O}(s^2)$ has the following general form:
\be
	\mathcal{M}(s)  \sim & \Delta \lambda  +  (\Delta\lambda)^2 \left( x_1 \log \frac{\mu^2}{s} +x_2 \right) \\
	&+ \left[ C^{(8)}_{Hi}  +  C^{(6)}_i \cdot C^{(6)}_j \left( y_1 \log \frac{\mu^2}{s} + y_2\right)  +  \Delta \lambda \cdot C_{Hi}^{(8)} \left( z_1 \log \frac{\mu^2}{s} +z_2\right)\right] \frac{s^2}{\Lambda^4},
	\label{eq:general}
\ee
where $\Delta \lambda$ is threshold correction to the SM coupling $\lambda$, and $(x_i,y_i,z_i)$ are numbers that need to be confirmed by loop calculations. The $C^{(6)}_i$'s are operators listed in Eq.(\ref{eq:dim6-H4}) and Eq.(\ref{eq:dim6-Hpsi}). {In Eq.\eqref{eq:general}, the $C^{(6)}\cdot \lambda$ terms disappear because such terms only linearly depend on  $s$, therefore can't enter the $d^2\mc{M}/ds^2$. }


{We note that, thought the energy dependence of terms in the first line of Eq.(\ref{eq:general}) is shown as $s^0$, these terms can contribute to $d^2\mc{M}/ds^2|_{s\to 0}$ too, either by introducing the mass regulator~\cite{Bellazzini:2020cot,Bellazzini:2021oaj,Arkani-Hamed:2020blm,Chen:2022nym} or defining generalized couplings~\cite{Arkani-Hamed:2020blm}. In the present paper, after the subtraction of the amplitude Eq.\eqref{eq:general}, log-terms in the first line also enter positivity bounds.}
For example, as we have indicated, the $\log{(\frac{\mu^2}{s})}$ will turn to be $\log{(\frac{\mu^2}{ \Lambda^2+s})}$ after we adopt the subtraction. The latter one in the neighbourhood of $s = 0$ is $\log{\frac{\mu^2}{ \Lambda^2}}-\frac{s}{ \Lambda^2}+\frac{s^2}{2 \Lambda^4}+\mc{O}(s^3)$, which can produce an $s^2$ term. The dim-8 operators together with the dim-4 and dim-6 operators can contribute to the  amplitude space $(\bl{m}_1,\bl{m}_2,\bl{m}_3)$ though loop diagrams in Fig.~\ref{fig:EFTdiagrams}. According to the completeness of space, the result can be generally expanded as
\bea
\bs{f}' &=& C'_1\ \bl{m}_1  + C'_2\ \bl{m}_2 + C'_3\ \bl{m}_3\;.
\eea
We evaluate diagrams in Fig.~\ref{fig:EFTdiagrams} and give $(C'_1,C'_2,C'_3)$ at $\mu= \Lambda$ as follows:
\bea
C'_1 = & C^{(8)}_{H1}  & -\frac{\Delta\lambda(36 C^{(8)}_{H1} +13 C^{(8) }_{H2}  +13 C^{(8) }_{H3} -18 \Delta \lambda )}{72 \pi ^2} \nonumber \\
&&+\frac{ 13 {C^{(6)2}_{H1}} + 26 {C^{(6)}_{H1}} {C^{(6)}_{H2}}+8 {C^{(6)2}_{H2}}- 10 {C^{(1) 2}_{H\psi_L}} +10 {C^{(3)2}_{H\psi_L}}-5 C_{H\psi_R}^2+5 C_{Hud}^2}{36 \pi ^2} \;, \nonumber \\
C'_2 = & C^{(8)}_{H2} & -\frac{\Delta\lambda(13 C^{(8) }_{H1} +36 C^{(8) }_{H2} +13 C^{(8) }_{H3} -18 \Delta \lambda )}{72 \pi ^2}\nonumber \\
&&+\frac{52 {C^{(6)2}_{H1}}-52{C^{(6)}_{H1}} {C^{(6)}_{H2}}+17 {C^{(6)2}_{H2}}+40 {C^{(1)2}_{H\psi_L}}+40 {C^{(3)2}_{H\psi_L}}+20 C_{H\psi_R}^2}{144\pi ^2} \;, \nonumber \\
C'_3 = & C^{(8)}_{H3} & -\frac{\Delta\lambda (75 C^{(8) }_{H1}+52 C^{(8) }_{H2}+121 C^{(8) }_{H3} -72 \Delta \lambda)}{72 \pi ^2}\nonumber \\
&&+\frac{56{C^{(6)}_{H1}}^2-26 {C^{(6)}_{H1}} {C^{(6)}_{H2}}-11{C^{(6)2}_{H2}}-40 C^{(3)2}_{H\psi_L}-10 C_{Hud}^2}{72 \pi ^2}\;,
\label{eq:loopC}
\eea
where for simplicity, we have assumed that there are only one generation and one color for fermions.  In the amplitude space, the cone structure determined by generators in Eq.(\ref{eq:generators}) remains unchanged. It's normal vectors still lead to the same forms of inequalities:
\bea
C'_2\geq0,\quad C'_1+C'_2\geq0,\quad C'_1+C'_2+C'_3\geq0\;,
\label{eq:loopbounds}
\eea
Unlike Eq.(\ref{eq:tree-bound}), the $C'_i$ has been modified by dim-4 and dim-6 coefficients as shown in Eq.(\ref{eq:loopC}), therefore inequalities in Eq.(\ref{eq:loopbounds}) are new positivity bounds at loop level. 

From a top-down perspective, starting from an UV theory, Wilson coefficients can be obtained by the matching. At the one loop level, the $C_{Hi}^{(8)}=C_{Hi}^{(8)\rm tree}+C_{Hi}^{(8)\rm loop}$ in Eq.(\ref{eq:loopC}), where $C_{Hi}^{(8) \rm tree }$'s are the tree-matching results, and the $C_{Hi}^{(8)\rm loop}$'s are obtained through the loop-matching conditions for various UV models. Nevertheless, it's enough to substitute rest coefficients in Eq.(\ref{eq:loopC}) into their tree-matching values since the coefficients are already suppressed by $\frac{1}{16\pi^2}$. As a result, if we keep Eq.(\ref{eq:loopC}) only in tree order, neglecting the terms suppressed by factor $\frac{1}{16\pi^2}$ will bring Eq.(\ref{eq:loopbounds}) back to the classical tree level positivity bounds. 
In principle, complete one-loop calculations should include those contributions from SM coupling $g_{\rm SM}^{}$. For example, let $g^{}_X$ be the coupling of the heavy particle $X$, contributions of order $\mc{O}(g_{\rm SM}^{} g_X^{})$ to $C'_i$'s exist too. Taking into account these contributions, the final expressions of $C'_i$'s will be more complex, 
but that don't change the basic idea of convex cone approach. However, including terms of order $\mc{O}(g_{\rm SM}g_X)$ into the positivity bounds is beyond the scope of this paper.
\subsection{Discussions}
Some discussions for our results are helpful. First of all, the calculation of the loop diagrams in Fig.~\ref{fig:EFTdiagrams} usually contains terms proportional to $s^2\log{(\frac{\mu^2}{s})}$. But, after the implementation of the subtraction scheme, such a log  becomes $s^2\log{(\frac{\mu^2}{\Lambda^2})}+\mc{O}(s^3)$, which provides contributions proportional to $\log{(\frac{\mu^2}{\Lambda^2})}$ to $C_i^\prime$'s. These contributions have been neglected in Eq.(\ref{eq:loopC}) by setting  $\mu=\Lambda$. However, the $\mu$-dependent terms with $\log{(\frac{\mu^2}{\Lambda^2})}$ play important roles when the RG running effects on positivity bounds are considered. The RG effects are discussed in Section~\ref{sec:RGE}.

To test the correctness of the loop level positivity bounds, the top-down perspective will be helpful. Firstly, we find that if the heavy particle $X$ is a fermion, then $X$ can't be the tree level completion of dim-8 operators $\mc{O}_{{Hi}}^{(8)}$'s, which means that $C_{Hi}^{(8) \rm tree}=0$. Except for the loop-matching coefficients $C_{Hi}^{(8) \rm loop}$, only the four $\psi^2 H^2 D$-type operators in Eq.(\ref{eq:dim6-Hpsi}) are possible to contribute to the amplitude $f'$ through diagrams of fermion loop. Focusing only on $\psi^2 H^2 D$-type coefficients in Eq.(\ref{eq:loopC}), we find that positivity bounds are always satisfied if the $X$ is a fermion:
\bea
C'_2&=&\frac{1}{36 \pi ^2}\left[10{C^{(1)2}_{H\psi_L}}+10{C^{(3)2}_{H\psi_L}}+5C_{H\psi_R}^2 \right]\geq0,\nonumber \\
C'_1+C'_2&=&\frac{1}{36 \pi ^2} \left[20{C^{(3)2}_{H\psi_L}}+5C_{Hud}^2 \right]\geq0,\nonumber \\
 C'_1+C'_2+C'_3&=& 0 \;.
 \label{eq:fermionC}
\eea

If the heavy particle is not a fermion, but still lead to most of tree level coefficients on r.h.s of Eq.(\ref{eq:loopC}) being zero, i.e. $\Delta \lambda=0$ and $C_{i}^{(n)\rm tree}=0$ for $n=6,8$, then only loop-matching coefficients $C_{Hi}^{(8)\rm loop}$'s will survive in Eq.(\ref{eq:loopC}). Hence, the positivity bounds in Eq.(\ref{eq:loopbounds}) must directly constrain $C'_i=C_{Hi}^{(8)\rm loop}$. Some examples have illustrated such a situation, such as adding a  heavy doublet with $Y = 1/2$ to the SM,  or the SM is extended by  heavy quadruplet with $Y = 1/2$ or with $Y = 3/2$ ~\cite{Chala:2021wpj}. In these three examples, as we have expected, the one-loop matching conditions do produce a set of vectors $\vec{C}^{(8)\rm loop}$ that satisfy the original tree level positivity bounds in Eq.(\ref{eq:tree-bound}).

However, if the tree level matching coefficients don't all vanish, it is difficult to see whether Eq.(\ref{eq:loopbounds}) must be met. Thereby we should study a specific UV model and match it onto the SMEFT in the $g_{\rm SM}\to 0$ limit, to obtain the accurate values for Wilson coefficients both at tree level and loop level. Typically, if tree level coefficients $C_{Hi}^{(8)}$'s satisfy original bounds in Eq.(\ref{eq:tree-bound}) and all bounds are greater than zero, loop modifications are generally smaller than tree level coefficients, thus will not alter positiveness of bounds. However, if one of the bounds in Eq.(\ref{eq:tree-bound}) equals to zero, this bound will be dominated by $C_{Hi}^{(8)\rm loop}$ and $C_{Hi}^{(6)}$, it's positiveness can't be guaranteed. In such a situation, several examples are shown in the next subsection.

\subsection{Examples}
Some extensions of the SM  have been studied for verifying the correctness of the positivity bounds. For example, we extend SM by a heavy charged triplet scalar $\Delta^I$ with $Y=-1$, which is the type-II seesaw model~\cite{Konetschny:1977bn,Magg:1980ut,Schechter:1980gr,Cheng:1980qt,Mohapatra:1980yp,Lazarides:1980nt}. This heavy particle has brought us a non-zero vector $\vec{C}\equiv(C_{H1}^{(8)},C_{H2}^{(8)},C_{H3}^{(8)})=g_{\Delta}^2(0,8,0)$ at the tree level. The tree-level positivity bounds are all greater than zero:
\bea
C_{H2}^{(8)}=8g_{\Delta}^2>0,\quad C_{H1}^{(8)}+C_{H2}^{(8)}=8g_{\Delta}^2>0, \quad C_{H1}^{(8)}+C_{H2}^{(8)}+C_{H3}^{(8)}=8g_{\Delta}^2>0\;.
\eea
Even if we add new contributions to $\vec{C}$ from the loop level, the bounds are just slightly disturbed by the loop level coefficients, and will not be destroyed.
For the same reason, most of the UV models  will not break the positivity bounds at loop level. However, as Ref.~\cite{Chala:2021wpj} has pointed out, by the loop-matching procedure two special models may produce Wilson coefficients that violate tree level bounds in Eq.(\ref{eq:tree-bound}). So it is worthwhile to study these models again under the new bounds in this paper.

\subsubsection*{heavy neutral scalar singlet + triplet}
We add a  heavy neutral scalar singlet $\mc{S}$ and a heavy neutral scalar triplet $\Xi$ to SM at the same time, the lagrangian of interaction is written as
\be 
\mc{L}_{\rm int}=g_{\mc{S}} M_{\mc{S}} \mc{S} H^\dag H+g_{\Xi} M_\Xi H^\dag \Xi^I \sigma^I H\;.
\ee
Letting $g_{\mc{S}}\to 0$ or $g_\Xi\to 0$ converts the lagrangian to the case of one-particle extension of SM. It is reasonable to include two heavy particles because there usually can be more than one type of heavy particle existing in the loop. Sometimes one-particle extension of the SM may not be enough. For simplicity, we assume the heavy particles have same masses: $M_{\mc{S}}=M_\Xi=\Lambda$. The tree level matching gives  following coefficients:
\begin{alignat}{3}
	\Delta \lambda &= \frac{-g_{\mc{S}}^2-g_\Xi^2}{2}, \quad &
	C_{H1}^{(6)\rm tree}&=\frac{g_{\mc{S}}^2-g_\Xi^2}{2},\quad &
	C_{H2}^{(6)\rm tree}&=-2g_\Xi^2,
	\nonumber \\
	C_{H1}^{(8)\rm tree}&=4g_\Xi^2,
	\quad & 
	C_{H2}^{(8)\rm tree}& =0,
	\quad & 
	C_{H3}^{(8)\rm tree}&=2g_{\mc{S}}^2-2g_\Xi^2.
	\label{eq:example1tree}
\end{alignat}
In the meanwhile, the loop-matching dim-8 coefficients are (see appendix~\ref{appB} for details)
\bea
C_{H1}^{(8)\rm loop}&=&-\frac{195g_{\mc{S}}^4-132g_{\mc{S}}^2 g_\Xi^2+535g_\Xi^4}{720 \pi^2}, \nonumber \\
	C_{H2}^{(8)\rm loop}&=&-\frac{39g_{\mc{S}}^4+78g_{\mc{S}}^2 g_\Xi^2+61g_\Xi^4}{144 \pi^2},
	\nonumber \\
	C_{H3}^{(8)\rm loop}&=&-\frac{187g_{\mc{S}}^4+1386g_{\mc{S}}^2 g_\Xi^2+271g_\Xi^4}{720 \pi^2}\;,
	\label{eq:matchingC1}
\eea
in which, as we can see, $C_{H2}^{(8)\rm loop}$ is negative. It obviously destroys one of the tree level bounds in Eq.(\ref{eq:tree-bound}). However, for our improved bounds in Eq.(\ref{eq:loopC}), we also need to substitute the tree-matching coefficients in Eq.(\ref{eq:example1tree}) into new bounds
\bea
 C'_2&=& \frac{3g_{\mc{S}}^4+6g_{\mc{S}}^2 g_\Xi^2+g_\Xi^4}{48 \pi^2}\geq0 \;\nonumber \\
 C'_1+C'_2&=& \frac{45 g_{\mc{S}}^4+191g_{\mc{S}}^2 g_\Xi^2+670g_\Xi^4}{360 \pi^2}\geq0 \; ,\nonumber \\
 C'_1+C'_2+C'_3&=&\frac{1433g_{\mc{S}}^4+836g_{\mc{S}}^2 g_\Xi^2+979g_\Xi^4}{720 \pi^2}\geq0 \;, 
 \eea
which are also true for $g_{\mc{S}}\to 0$ or $g_\Xi\to 0$. In each limit, the loop-matching coefficients in Eq.(\ref{eq:matchingC1}) can reproduce the results of Ref.~\cite{Chala:2021wpj}. This example has shown that when SM is extended by neutral scalar singlet or heavy neutral scalar triplet, even if the more than one types of heavy particles run in the loop, our new positivity bounds are valid.

\subsubsection*{The type-I seesaw model}
The type-I seesaw model naturally explains the tiny masses of neutrino through the seesaw mechanism~\cite{Minkowski:1977sc, Yanagida:1979as, Gell-Mann:1979vob, Glashow:1979nm, Mohapatra:1979ia}. This model introduces a heavy right-hand fermion singlet $N_{\rm R}$ to the SM.
As we have mentioned, the case is particularly interesting when the UV particle is a fermion, because the fermion can only lead to a zero vector of $\vec{C}_H^{(8)}$ at the tree level. The coefficient $C_{Hi}^{(8)}$ receives contribution through the fermion loop. We find that
$N_{\rm R}$ has a Yukawa interaction with the left-hand lepton doublet $\ell_{\rm L}$: 
\begin{equation}
	\mathcal{L}_{\mathrm{int}}=-\overline{\ell_{\mathrm{L}}} Y_\nu \widetilde{H} N_{\mathrm{R}}+\text {h.c.}\;.
\end{equation}
The tree level matching of the type-I seesaw is well known, and it produces the seesaw effective field theory (SEFT)~\cite{Zhang:2021tsq}. SEFT-I contains a dim-5 Weinberg operator and a dim-6 operator $
\mathcal{O}^{(6)}_{\rm SEFT-I}=(\overline{\ell_{ \mathrm{L}}} \tilde{H}) \mathrm{i} \slashed{\partial} (\tilde{H}^{\dagger} \ell_{ \mathrm{L}})$. The Weinberg operator generates Majorana masses of neutrinos after the spontaneous gauge symmetry breaking. The dim-6 operator will cause the unitarity violation of the lepton flavor mixing matrix through modifying the normalisation of $\ell_{\rm L}$. The dim-6 operator should be rewritten in the Warsaw basis~\cite{Grzadkowski:2010es}:
\begin{equation}
	\mathcal{O}^{(6)}_{\rm SEFT}=\frac{1}{4}\left[\left(\overline{\ell_{  \mathrm{L}}} \gamma^\mu \ell_{ \mathrm{L}}\right)(H^{\dagger} i \overleftrightarrow{D}_\mu H )-\left(\overline{\ell_{  \mathrm{L}}} \gamma^\mu \sigma^I \ell_{  \mathrm{L}}\right) (H^{\dagger} i \overleftrightarrow{D}_\mu^I H )\right]\;,
\end{equation}
where the flavor index has been dropped since for simplicity only one generation is considered. Hence, the tree level matching coefficients that are relevant for positivity bounds are simple:
\begin{alignat}{3}
	\Delta \lambda &= 0, \quad &
	C_{H1}^{(6)\rm tree}&=0,\quad &
	C_{H2}^{(6)\rm tree}&=0,
	\nonumber \\
	C_{H1}^{(8)\rm tree}&=0,
	\quad & 
	C_{H2}^{(8)\rm tree}& =0,
	\quad & 
	C_{H3}^{(8)\rm tree}&=0. \nonumber\\
	C^{(1)}_{H\psi_L} &= \frac{Y_\nu^2}{4}, \quad &
	C^{(3)}_{H\psi_L} &= -\frac{Y_\nu^2}{4}.
\end{alignat}
The one-loop matching values of $C_{Hi}^{(8)}$'s are obtained by evaluating fermion loops:
\bea
C_{H1}^{(8)\rm loop}&=&\frac{Y_\nu^4}{144\pi^2},\quad  
	C_{H2}^{(8)\rm loop}=\frac{59Y_\nu^4}{288\pi^2},\quad
	C_{H3}^{(8)\rm loop}=\frac{Y_\nu^4}{30\pi^2}\;.
\eea
These coefficients are all positive so tree level positivity bounds are naturally satisfied. In addition, adding the contributions from $C^{(1)}_{H\psi_L}$ and $C^{(3)}_{H\psi_L}$ does not alter the positiveness of bounds because of Eq.(\ref{eq:fermionC}). More examples for fermion loop contributions have been considered in Ref.~\cite{Zhang:2021eeo}, in which the fermion loops are studied by the Cutkosky cutting rules~\cite{Cutkosky:1960sp}. The amplitudes from those examples are still constrained inside the convex cone, thus are allowed by positivity bounds. Under these conditions, adding the dim-6 contributions will only strengthen the bounds, as shown in Eq.(\ref{eq:fermionC}).
\section{RGE effect}
\label{sec:RGE}
In this section, we will focus on the RG running effect on the positivity bounds, i.e. explore the form of bounds at the scale $\mu<\Lambda$. For the consistency of one-loop calculations, as $\mu$ decreases, only $C_{Hi}^{(8)}$ will evolve with the scale according to it's one-loop beta function. In Eq.(\ref{eq:loopC}), we don't need  to consider the RG running of dim-4 and dim-6 coefficients since they have been suppressed by the loop factor $\frac{1}{16\pi^2}$. 
In order to derive the one-loop beta function of $C_{Hi}^{(8)}$, we need to calculate the divergences of all the diagrams in Fig.~\ref{fig:EFTdiagrams}. The exact forms of beta functions for dim-8 bosonic operators are given in Refs.~\cite{Chala:2021pll,DasBakshi:2022mwk}. 

In addition to the running of $C_{Hi}^{(8)}$ with the scale, we also need to consider the changes of bounds themselves with the scale. Note, we have mentioned that Eq.(\ref{eq:loopC}) is obtained at $\mu= \Lambda$, so the $\log{(\frac{\mu^2}{ \Lambda^2})}$-dependent terms have been dropped. In consideration of the RG effect, the log-dependent terms must be put back explicitly
\bea
\left. C'_1 \right|_{\rm log} &=& \frac{16{C_1^{\left(6\right)2}}+32C_1^{(6)}C_2^{\left(6\right)}+11{C_2^{\left(6\right)2}}-16({C_{H\psi_L}^{(1)2}}-{C_{H\psi_L}^{(3)2}})-8({C_{H\psi_R}^2}-{C_{Hud}^2})}{96\pi^2}  \log{\left(\frac{\mu^2}{ \Lambda^2}\right)} \;, \nonumber \\ 
\left. C'_2 \right|_{\rm log} &=& \frac{16{C_1^{\left(6\right)2}}-16C_1^{\left(6\right)}C_2^{\left(6\right)}+5{C_2^{\left(6\right)2}}+8 (2{C_{H\psi_L}^{\left(1\right)2}}+2{C_{H\psi_L}^{\left(3\right)2}}+{C_{H\psi_R}^2} )}{96\pi^2} \log{\left(\frac{\mu^2}{ \Lambda^2}\right)} \;, \nonumber \\
\left. C'_3 \right|_{\rm log} 
&=& \frac{40{C_1^{\left(6\right)2}}-16C_1^{\left(6\right)}C_2^{\left(6\right)}-7{C_2^{\left(6\right)2}}-8(4{C_{H\psi_L}^{\left(3\right)2}}+{C_{Hud}^2})\ }{96\pi^2} \log{\left(\frac{\mu^2}{ \Lambda^2}\right)}\;.  
\label{eq:log-C}
\eea
These log-dependent terms also come from diagrams in Fig.~\ref{fig:EFTdiagrams}, more explicitly, they are the $\log{(\frac{\mu^2}{s})}$'s appearing in the loop calculation.  The subtraction scheme turns $\log{(\frac{\mu^2}{s})}$ to $\log{(\frac{\mu^2}{ \Lambda^2})}+\mc{O}(s)$. Collecting the $\log{(\frac{\mu^2}{ \Lambda^2})}$ terms brings us Eq.(\ref{eq:log-C}).

We find that Eq.(\ref{eq:log-C}) has exactly the same forms of the beta functions of $C_{Hi}^{(8)}$'s in Ref.~\cite{Chala:2021pll} up to an overall minus sign. This observation means that \emph{the RG evolution of  $C_{Hi}^{(8)}$ is just offset by the change of the positivity bounds with the scale}. Therefore, once the positivity bounds are satisfied at the scale $\mu=\Lambda$, they will be valid at any scale {below $\Lambda$}.

This is a general conclusion at one-loop level. Because when we calculate the beta functions of $C_{Hi}^{(8)}$'s and consider the $\mu$-dependence of bounds in Eq.(\ref{eq:log-C}), we are implementing the evaluation of same loop diagrams in EFT, such as Fig.\ref{fig:EFTdiagrams}. 
{To illustrate, we firstly denote the general expression for the running of dim-8 coefficients as: $16\pi^2\mu \frac{dC^{(8)}_i}{d\mu}=\gamma_{ij} C_j^{(8)} + \gamma'_{ijk} C^{(6)}_j C^{(6)}_k $. This expression has taken into account contributions from loops involving single insertions of dimension-eight operators as well as from pairs of dimension-six operators. 
To determine the $\gamma$ for $C_{Hi}^{(8)}$, we calculate loops of 4-$H$ processes and give exactly the same amplitude given in Eq.\eqref{eq:general}($g_{\rm SM} \to 0$ is taken too). After that, requiring $\mc{M}(s)$ is scale-independent provides that $\gamma=-32\pi^2 z_1$ and $\gamma'=-32\pi^2 y_1$, where $y_1$'s and $z_1$'s are given in Eq.\eqref{eq:log-C}. Finally, let's consider the effective coupling $C'$ at an arbitrary scale $\mu$ that is below $\Lambda$: 
\bea
C'(\mu)=C^{(8)}_{H}(\Lambda)+\left(\frac{\gamma}{32\pi^2}+z_1\right)\log(\frac{\mu^2}{\Lambda^2})\cdot C^{(8)}_{H}+C^{(6)}\cdot\left(\frac{\gamma'}{32\pi^2}+y_1\right)\log(\frac{\mu^2}{\Lambda^2}) \cdot C^{(6)}\;,
\label{eq:runningC} 
\eea
where the second term and the third term vanish. Eventually, effective couolings are free of $\log(\mu)$, and positivity bounds are scale-independent.
}

To conclude, the RG running will not bring new physical information to positivity bounds.
When we can actually measure the values of Wilson coefficients at low-energy experiments, the coefficients are determined at the characteristic scale of energy. To test the positivity, the bounds need to be modified by adding Eq.(\ref{eq:log-C}), and are expected to be valid at this characteristic scale.

\section{Summary}
\label{sec:summary}
In present paper, we have used the framework of convex geometry to explore the general formula for the leading $s^2$ positivity bounds. The formula can be applied to both the tree level and the loop level. As an explicit application, we study  positivity bounds on the 4-$H$ forward scattering at the loop level.  In order to remove singularities of amplitude in the IR-region without losing UV information, we have defined the subtracted amplitude. In this way, we can well handle the singularities at $s=0$ of poles and the $\log{(\frac{\mu^2}{s})} $-function appearing in loop calculations of EFT. 

The advantage of the convex cone framework is that it uses the symmetry information of scattering particles. The matrix $f'_{ijkl}$ in the ``amplitude space'' has a universal structure determined by bases of the space, regardless of whether the $f'_{ijkl}$ is calculated by the tree diagram or the loop diagram. As long as the completeness of the dim-8 effective operator set in SMEFT~\cite{Li:2020gnx,Murphy:2020rsh} is ensured, we can treat the characteristic amplitude matrices of these effective operators as bases of the amplitude space. Any $f'_{ijkl}$ can be expanded under this group of bases. At the same time, we can calculate the generators of convex cone with CG coefficients. With the generators, the structure of the convex cone is known in the amplitude space, which will not be altered by the loop calculation. What needs to be changed is the way how the loop amplitude enters the amplitude space. The amplitude is constrained inside the convex cone, whose normal vectors correspond to positivity bounds. To conclude, positivity bounds (whether with or without the subtraction) are always the same. But when they are explained in terms of specific Wilson coefficients, they will change if loop diagrams are added.

We then use the convex cone approach to reproduce the tree level positivity bounds for 4-$H$ process, and further study the effect of loop diagrams in EFT on positivity bounds. 
Generally, the dim-4 and dim-6 operators will also get into the loop diagrams, hence their coefficients contribute to the amplitude space. We have obtained new positivity bounds at loop level, which have similar form to the tree level bounds. These loop bounds include small modifications to the original tree bounds. The loop modifications are all suppressed by the loop factor $\frac{1}{16\pi^2}$, and are induced by the dim-4 and the dim-6 operators. Neglecting these terms will bring the loop bounds back to classical tree level result. In addition, we perform the matching calculation for several special UV models to check whether the obtained loop bounds are valid. These models include the heavy neutral scalar singlet, heavy neutral scalar triplet, the type-I seesaw model and type-II seesaw model.

Finally, we investigate the influence of RGE on the positivity. Our result have shown that RG effect will not cause any violation of bounds. Besides, as long as bounds are satisfied at the cutoff (or matching) scale $\Lambda$, they will also be satisfied at any scale {below $\Lambda$}. This is because the new loop bounds change with the scale, and this change is just offset by the RG running of the dim-8 coefficient. 
The underlying physical reason is that the logarithmic terms of these two evolutions come from same loop diagrams and will cancel each other.

This work is the first to study the loop effect of leading $s^2$ positivity bounds in the SMEFT. Although only the forward scattering of 4-$H$ is considered, we have pointed out that the theoretical framework can be extended to other cases, such as the scattering of vector bosons or fermions. The loop effects on the bounds of these processes have not been effectively understood, and we hope to come back to these issues in the near future.

\section*{Acknowledgments}
The author would like to thank Profs. Shun Zhou, Shuang-Yong Zhou for helpful discussions and their valuable suggestions. And the author especially thank Mikael Chala for the discussion. This work was supported in part  by the National Natural Science Foundation of China under grant No. 11835013 and the Key Research Program of the Chinese Academy of Sciences under grant No. XDPB15.

\appendix
\section{Example of the subtraction scheme}
\label{appA}
As simple example, this appendix discusses how to subtract the singularities in low energy region from a given amplitude. Suppose that an amplitude has singularities that include a low energy pole, a high energy pole, and a branch cut from 0 to $\infty$, namely,
\bea
M\left(s\right)=\frac{1}{s-m^2}+\frac{1}{s-M^2}+\log{\left(-s\right)}
\eea
with $m^2 < \Lambda^2< M^2$.

\begin{figure}[t]
	\centering  
	\includegraphics[width=0.45\linewidth]{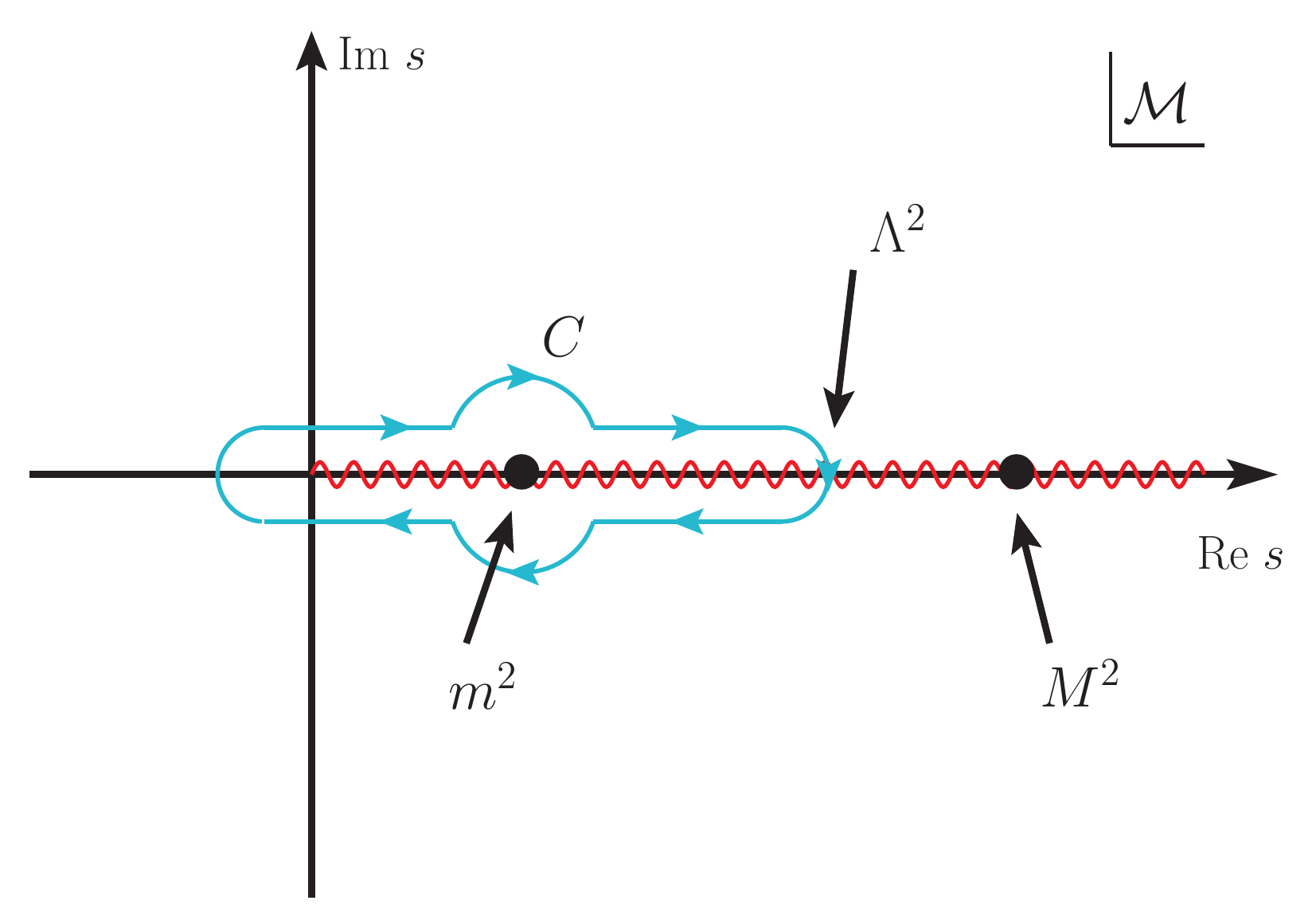}
	\caption{The contour for the amplitude in the low-energy region.}
	\label{fig:contour-3}
\end{figure}

To subtract the low-energy singularity, we consider the contribution of an integral from the low energy region ($0<s<\Lambda^2$). The contour is shown in Fig.~\ref{fig:contour-3}, where black points correspond to poles and  the red cut corresponds to the branch cut of the log function. The contour contains a small circle which picks up the pole at $s^\prime=m^2$, and an integral on discontinuity of the branch cut with interval from 0 to $\Lambda^2$. The low-energy amplitude is given by 
\bea
M_{\rm low}=\frac{1}{2\pi i} \left[ \oint_{C}{ds^\prime\frac{M\left(s^\prime\right)}{s^\prime-s}} + \int_0^{\Lambda^2}{\frac{ {\rm Disc} M\left(s^\prime\right)}{s^\prime-s}ds^\prime} \right]
=\frac{-
1}{m^2-s}-\left.\log{\left(s^\prime-s\right)}\right|_0^{\Lambda^2}
\eea
where the ${\rm Disc} \log{(-s)}=- 2\pi i$ has been used. Therefore, we define the subtracted amplitude as 
\bea
M'\left(s\right)\equiv M\left(s\right)-M_{\rm low}=\frac{1}{s-M^2}+\log{\left(\Lambda^2-s\right)}
\eea
We can see that now the branch cut is starting from $s=\Lambda^2$, low-energy pole $s=m^2$ has been dropped from the subtracted amplitude, but the high-energy pole $s=M^2$ is left. Hence, the subtracted amplitude is totally analytical in low-energy region but it's UV information is still kept. 

\section{Details of the matching procedure}
\label{appB}
We calculate two kinds of one-loop diagrams in the UV theory to obtain the one-loop matching coefficients, i.e. the one-light-particle-irreducible (1LPI) diagrams and diagrams with corrected external legs. The on-shell forward amplitude of the 1LPI diagrams are evaluated  in the hard momentum region. In addition, the loop in the external leg produces a threshold correction to Higgs's kinetic term, which contributes to dim-8 coefficients through equation of motion. Summing these two kinds of diagrams up and matching the results to the dim-8 operator bases, one-loop matching coefficients are acquired.

In our work, we  evaluate the diagrams with the help of some semi-automatic programs in $\mathtt{Mathematica}$. Firstly, we use $\mathtt{FeynRules}$ ~\cite{Christensen:2008py} to generate the $\mathtt{FeynArts}$ ~\cite{Hahn:2000kx} model files for various models. The interface $\mathtt{FeynHelper}$ ~\cite{Shtabovenko:2016whf} is utilized to realize the connection among $\mathtt{FeynArts}$, $\mathtt{FeynCalc}$ ~\cite{Shtabovenko:2016sxi} and $\mathtt{Package{\text -}X}$ ~\cite{Patel:2015tea}. More explicitly, $\mathtt{FeynArts}$ will provide all the Feynman diagrams and corresponding amplitudes. After that, loop integrals in the amplitudes are calculated with the help of $\mathtt{FeynCalc}$. Finally, analytical expressions for Passarino-Veltman functions ~\cite{Passarino:1978jh} are automatically provided by $\mathtt{Package{\text -}X}$.


\bibliographystyle{JHEP}
\bibliography{refs}

\end{document}